\documentclass[12pt,letterpaper]{article}
\pdfoutput=1

\usepackage{amsmath,amssymb,calc, amsthm}
\usepackage{graphicx}
\usepackage{hyperref}
\usepackage[nosort]{cite}
\usepackage{epsfig,psfrag}

\makeatletter
\renewcommand\section{\@startsection {section}{1}{\z@}%
                                   {-3.5ex \@plus -1ex \@minus -.2ex}
                                   {2.3ex \@plus.2ex}%
                                   {\normalfont\large\bfseries}}
\renewcommand\subsection{\@startsection{subsection}{2}{\z@}%
                                     {-3.25ex\@plus -1ex \@minus -.2ex}%
                                     {1.5ex \@plus .2ex}%
                                     {\normalfont\bfseries}}
\makeatother

\theoremstyle{plain}

\theoremstyle{definition}

\let\non\nonumber

\def\one{^{(1)}}

\newcommand{\bea}{\begin{eqnarray}}
\newcommand{\eea}{\end{eqnarray}}
\newcommand{\be}{\begin{equation}}
\newcommand{\ee}{\end{equation}}
\newcommand{\bma}{\begin{pmatrix}}
\newcommand{\ema}{\end{pmatrix}}

\newcommand{\Z}{{\mathbb Z}}
\newcommand{\R}{{\mathbb R}}

\newcommand{\PP}{{\mathbb P}}

\newcommand{\p}{\partial}

\newcommand{\C}[1]{$(\ref{#1})$}

\def\IZ{\relax\ifmmode\mathchoice
{\hbox{\cmss Z\kern-.4em Z}}{\hbox{\cmss Z\kern-.4em Z}}
{\lower.9pt\hbox{\cmsss Z\kern-.4em Z}} {\lower1.2pt\hbox{\cmsss
Z\kern-.4em Z}}\else{\cmss Z\kern-.4em Z}\fi}
\def\IR{\relax{\rm I\kern-.18em R}}

\def\one{{\hbox{ 1\kern-.8mm l}}}

\def\tr{{\rm tr\,}}

\newlength{\bredde}
\def\slash#1{\settowidth{\bredde}{$#1$}\ifmmode\,\raisebox{.15ex}{/}
\hspace*{-\bredde} #1\else$\,\raisebox{.15ex}{/}\hspace*{-\bredde}
#1$\fi}

\newsavebox{\zzzbar}
\sbox{\zzzbar}
  {\setlength{\unitlength}{0.9em}
  \begin{picture}(0.6,0.7)
  \thinlines
  \put(0,0){\line(1,0){0.6}}
  \put(0,0.75){\line(1,0){0.575}}
  \multiput(0,0)(0.0125,0.025){30}{\rule{0.3pt}{0.3pt}}
  \multiput(0.2,0)(0.0125,0.025){30}{\rule{0.3pt}{0.3pt}}
  \put(0,0.75){\line(0,-1){0.15}}
  \put(0.015,0.75){\line(0,-1){0.1}}
  \put(0.03,0.75){\line(0,-1){0.075}}
  \put(0.045,0.75){\line(0,-1){0.05}}
  \put(0.05,0.75){\line(0,-1){0.025}}
  \put(0.6,0){\line(0,1){0.15}}
  \put(0.585,0){\line(0,1){0.1}}
  \put(0.57,0){\line(0,1){0.075}}
  \put(0.555,0){\line(0,1){0.05}}
  \put(0.55,0){\line(0,1){0.025}}
  \end{picture}}

\newcommand{\ena}{\end{eqnarray}}
\newcommand{\beqa}{\begin{eqnarray}}
\newcommand{\eeqa}{\end{eqnarray}}

\def\bes #1\ees{\begin{split}#1\end{split}}

\newfont{\goth}{ygoth.tfm scaled 1200}                   

 \numberwithin{equation}{section}

\def\1{{(1)}}
\def\2{{(2)}}
\def\3{{(3)}}

\def\1{{\bf 1}}

\def\M{{\mathcal M}}

\def\1{{\bf 1}}
\def\3{{\bf 3}}
\def\7{{\bf 7}}
\def\2{{\bf 2}}
\def\8{{\bf 8}}

\def\cL{{\cal L}}
\def\dVol{\operatorname{dVol}}

\newcommand\Jb{\overline{J}}

\def\ep{{\epsilon}}

\def\pb{{\overline{\partial}}}

\def\bes #1\ees{\begin{split}#1\end{split}}

\begin{document}
\begin{titlepage}

\begin{center}

{March 31, 2015}
\hfill         \phantom{xxx}  EFI-15-11

\vskip 2 cm {\Large \bf Constraining de Sitter Space in String Theory} 
\vskip 1.25 cm {\bf  David Kutasov$^{a}$, Travis Maxfield$^{a}$, Ilarion Melnikov$^{b}$ and Savdeep Sethi$^{a}$}\non\\
\vskip 0.2 cm
 $^{a}${\it Enrico Fermi Institute, University of Chicago, Chicago, IL 60637, USA}
\vskip 0.2 cm
$^{b}${\it Department of Mathematics, Harvard University, Cambridge, MA 02138, USA}

\vskip 0.2 cm
{ Email:} \href{mailto:dkutasov@uchicago.edu}{dkutasov@uchicago.edu}, \href{mailto:maxfield@uchicago.edu}{maxfield@uchicago.edu}, \href{mailto:ilarion@math.harvard.edu}{ilarion@math.harvard.edu}, \href{mailto:sethi@uchicago.edu}{sethi@uchicago.edu}

\end{center}
\vskip 1 cm

\begin{abstract}
\baselineskip=18pt

We argue that the heterotic string  does not have classical vacua corresponding to de Sitter space-times of dimension four or higher. The same conclusion applies to type II vacua in the absence of RR fluxes. 
Our argument extends prior supergravity no-go results to regimes of high curvature.  We discuss the interpretation of the heterotic result from the perspective of dual  type II orientifold constructions. Our result suggests that the genericity arguments used in string landscape discussions should be viewed with caution.

\end{abstract}

\end{titlepage}


\section{Introduction} \label{intro}

The construction of accelerating space-times in string theory is a central open problem. We will focus on the case of maximally symmetric de Sitter space-times, which provide a good model for our  observed universe. The difficulty in realizing acceleration in string theory has simple origins. All classical sources of stress-energy in the low-energy effective theories that emerge from string theory or M-theory obey the strong energy condition (SEC). These sources, which include branes, anti-branes, fluxes and smooth choices of metric, cannot produce acceleration without explicit time-dependence in internal directions~\cite{Gibbons:1984kp, Maldacena:2000mw, Dasgupta:2014pma}.  

We are therefore left with the following question: can intrinsically stringy ingredients, like orientifold planes, perhaps in conjunction with quantum effects, lead to acceleration? If not, then the current paradigm for connecting string theory to the observed universe via compactification is in jeopardy. We might then imagine other starting points which can  evade the no-go constraints; for example, compactifications on spaces with boundaries, including the special case of localizing gravity on domain walls. 

The goal of this work is to strengthen prior no-go results to include stringy $(\alpha')$ effects. One of the difficulties with progress on the string landscape is that most proposed constructions involve poorly understood ingredients. We will evade many of these complications by studying a corner of string theory where classical stringy effects can be robustly analyzed, namely the heterotic string. Unlike type II theories, type I string theory or M-theory, the classical physics of flux compactifications can be studied in the heterotic string using currently available world-sheet technology. 

We will provide an argument based on symmetries that essentially rules out de Sitter space-times in the classical heterotic string.\footnote{The same conclusion applies directly to type II backgrounds without RR fluxes.}  The ten-dimensional space-time action of this theory has the schematic (string frame) form
\be \label{spacetimeaction}
S = {1\over 2 \kappa^2} \int d^{10}x \sqrt{-g} e^{-2\Phi} \left[ R + {\alpha'\over 4} \tr |R_+|^2 + {\alpha'^3} R^4+\ldots \right].
\ee
The omitted terms include an infinite set of higher curvature interactions involving the metric, dilaton $\Phi$, $B$-field, gauge-fields and fermions. We will argue that this action does not have de Sitter solutions  of dimension four or higher with a  cosmological constant that remains finite in the classical limit $g_s= e^{\Phi} \rightarrow 0$. 

From the world-sheet $\sigma$-model point of view, our analysis captures all perturbative $\alpha'$ corrections, non-perturbative effects like world-sheet instantons, as well as high curvature solutions, which cannot be described in an $\alpha'$ expansion.  It applies to both stable and unstable de Sitter space-times. Supersymmetry plays no role in our discussion.  

One might wonder whether the interactions captured in the tree-level action~\C{spacetimeaction}\ can ever give rise to space-time solutions with a cosmological constant. The answer to that question is certainly yes: $AdS_3 \times S^3 \times T^4$ is an exact tree-level solution of the heterotic string with suitable $H_3$-flux (see e.g.~\cite{Kutasov:1998zh}). In addition, the action~\C{spacetimeaction}\ certainly contains higher derivative interactions that violate the strong energy condition (SEC); in principle, solutions with a positive cosmological constant are possible. If anything, the absence of de Sitter solutions to~\C{spacetimeaction}\ would appear strange from the perspective of genericity arguments.     

The classical heterotic string involves many of the ingredients that are used in existing landscape constructions. We can ponder what ingredients are omitted and what are the implications of our results for such constructions. What we assume is the validity of the perturbative string starting point; namely, a world-sheet conformal field theory. The only ingredients not captured by our argument are quantum corrections controlled by the string coupling constant.  These include both loop effects and string non-perturbative effects. At least string non-perturbative effects play a significant role in many heterotic landscape proposals. For a recent discussion of heterotic landscape constructions based on Calabi-Yau compactifications, see~\cite{Anguelova:2010qd, Cicoli:2013rwa}.

Our world-sheet analysis both extends and complements the space-time analysis of~\cite{Green:2011cn, Gautason:2012tb}. Specifically, the space-time analysis of~\cite{Green:2011cn}\  applies to any accelerating solution of FLRW type rather than just de Sitter space, but it only accounts for the leading stringy $\alpha'$ effects. This analysis has recently been extended to include the effect of gaugino condensation~\cite{Quigley:2015jia}. On the other hand, the world-sheet argument given here applies specifically to the de Sitter case, but it captures all world-sheet effects. For prior interesting work discussing world-sheet constraints on acceleration, see~\cite{Schalm:2010hb, Parikh:2015bja}.

\subsubsection*{Branes and dual interpretations}

Two final comments are in order: the first concerns branes. Aside from fundamental strings, the heterotic string only contains NS5-branes and anti-NS5-branes.  In flat space, these branes can smoothly fatten into space-time gauge instantons and anti-instantons, which are perturbative configurations with a bounded string coupling. Indeed, there are many backgrounds that involve NS5-branes in which the string coupling is bounded. We expect our world-sheet argument to apply to all backgrounds in which the string coupling does not diverge. By restricting to backgrounds with bounded string coupling, our intuition is that we are missing no essential new ingredient provided by branes.  

The second comment concerns duality. Many type II and M-theory landscape constructions begin with a four-dimensional large volume string compactification preserving N=1 space-time supersymmetry. Supersymmetry is then spontaneously broken at a scale well below the Kaluza-Klein scale. For models with $F$-term supersymmetry breaking, generic quantum corrections to the space-time superpotential can, in principle, give de Sitter critical points; see, for example,~\cite{Westphal:2006tn}\ for a type IIB scenario.

One can imagine considering exactly this scenario with the perturbative heterotic string as the starting point, and one might be inclined to expect a qualitatively similar result of many de Sitter solutions. The heterotic space-time superpotential is corrected by world-sheet instantons, as well as by space-time instantons and one-loop in string perturbation theory. Our general result, which we stress requires no low-energy supersymmetry, captures the world-sheet instantons exactly along with all $\alpha'$ corrections to the space-time K\"ahler potential. With these ingredients, there are no macroscopic $dS_n$ minima for ${n>3}$.   

We can try to interpret this result in terms of dual descriptions. In general, it does not make sense to try to map loop and non-perturbative effects between duality frames. A perturbative expansion in one theory is usually not useful for learning about perturbative expansions in a dual description. The cases where duality is useful involve low-energy quantities like space-time superpotentials. These are precisely the ingredients that go into typical landscape constructions.

We can get a feel for what we are including and omitting in the language of type II orientifold constructions in the following way.  Let us recall the following chain of dualities: the $Spin(32)/\Z_2$ heterotic string is S-dual in ten dimensions to the type I string. The heterotic string on $T^2$ is dual to F-theory on a $K3$ surface. Similarly, the heterotic string on $T^3$ is dual to M-theory on a $K3$ surface, while the $T^4$ case is dual to type IIA on $K3$. 

These are exact dualities, which can be used to relate type I and heterotic backgrounds, and type II and M-theory backgrounds with a $K3$ factor to heterotic vacua with torus factors. While these are exact dualities, they are typically only useful for computing special quantities like the space-time superpotential. 

Let us take one of the most studied examples of a type IIB flux vacuum~\cite{Dasgupta:1999ss}; namely F-theory on  $K3\times K3$, which has an orientifold limit corresponding to type IIB on 
\be \M = {T^2\over \Omega (-1)^{F_L}\Z_2} \times K3. \label{specialcase}\ee 
One can turn on $G_3$-fluxes that preserve some or no supersymmetry while solving the leading order equations of motion. The breaking of supersymmetry by flux corresponds to $F$-term breaking in the four-dimensional effective supergravity description.

The basic non-perturbative effects that can renormalize the space-time superpotential correspond to Euclidean wrapped D3-branes in type IIB. These branes wrap divisors of $\M$. There are two basic cases from which one can construct the general divisor: the brane instanton can wrap $\left( {T^2/\Z_2} \sim\PP^1 \right) \times D_2$, where $D_2$ is a divisor of $K3$, or it can wrap the $K3$ surface of $\M$. The Euclidean D3-brane wrapping $\PP^1$ corresponds to the dual heterotic string, so the first class of corrections corresponds to world-sheet instantons captured by our argument. The case of wrapping the $K3$ surface corresponds to a wrapped NS5-brane in the heterotic dual description, which we do not capture. In the general case of type IIB on a $3$-fold $X_3$ with a $\PP^1$-fibration, the only brane instantons we miss are those that include the base of the fibration. 
 
Orientifold planes play a crucial role in the construction of type IIB flux vacua. Without orientifold planes, compact solutions with flux are not possible. Orientifold planes can both have negative tension and support SEC violating interactions; for this reason, it is commonly suggested that orientifolds might lead to the evasion of the supergravity no-go theorems prohibiting acceleration. The heterotic string has no orientifold planes. Yet the orientifold supported couplings that make possible type IIB flux vacua are already present in the classical heterotic string. An example of such a coupling is the gravitational modification of the heterotic Bianchi identity:
\be
dH_3 = {\alpha'\over 4}\left[ \tr(R\wedge R) - \tr(F\wedge F)\right]. 
\ee    
Along with this modification come assorted SEC violating interactions, but our result shows that the mere presence of these effects is not sufficient to generate de Sitter solutions.  We expect a qualitatively similar conclusion in a type II setting:  the mere presence of orientifold planes is not sufficient to generate de Sitter solutions.

\section{Space-time and World-sheet Symmetries}

The isometries of $dS_n$ and $AdS_n$ are most easily seen by realizing these space-times as submanifolds of $\R^{1, n}$ and $\R^{2,n-1}$, respectively, 
 \be
 \label{oonn}
 -x_0^2 \pm x_{n}^2+ \sum_{i=1}^{n-1} x_i^2 = \pm L^2,
 \ee 
 where the ambient space is endowed with the canonical metric:
 \be
 ds^2 = -dx_0^2 \pm dx_{n}^2+ \sum_{i=1}^{n-1} dx_i^2.
 \ee
The $+$ sign gives $dS$ while the $-$ sign gives $AdS$, and the length scale $L$ sets the radius of curvature of the space-time. The connected component of the isometry group of $dS_n$ is therefore the $SO(1,n)$ Lorentz symmetry group of the ambient Minkowski space-time. 

We are interested in the question of whether an $n$-dimensional de Sitter space-time can be realized in classical heterotic string theory. In general, heterotic string backgrounds are described by a world-sheet CFT with $(1,0)$  supersymmetry and central charge $(c,\overline{c}) = (15,26)$, tensored with the ghost sector of $(1,0)$ world-sheet supergravity. Since we want to go beyond the supergravity approximation, it is natural to ask what we mean by a de Sitter background. We will require the world-sheet CFT to have the following properties:
\renewcommand\labelenumi{(\theenumi)}\begin{enumerate}
\item The $SO(1,n)$ symmetry should be realized as an exact symmetry of the string theory. In particular, the world-sheet theory should have conserved currents $(J^a,\bar J^a)$, where $a$ runs over the adjoint representation of $SO(1,n)$, that satisfy the conservation equation 
\begin{align}
\label{ttww}
\pb J^a + \p \Jb^a = 0~.
\end{align}
The charge associated with \C{ttww},
\be
 Q^a=\oint dz J^a+\oint d\bar z\bar J^a~,
 \ee 
must be a physical observable in string theory. This means that the $J^a$ transform under the world-sheet superconformal group as top components of superfields whose bottom components are primaries of dimension $({1\over 2},0)$, and $\bar J^a$ are conformal primaries of dimension $(0,1)$. 

\item
We will further assume that the background can be Wick rotated to a Euclidean one by taking $x^0\to ix^0$ in~\C{oonn}, and that this gives a compact unitary CFT with symmetry $SO(n+1)$ realized again by currents satisfying~\C{ttww}. A Wick rotation of this sort is often  impossible. For example, the sigma model on the $SU(2)$ group manifold cannot be Wick rotated to give three-dimensional de Sitter space, due to the presence of a three-form field strength on the three-sphere, $H=dB$, which is proportional to the volume form on the sphere. This  field strength is real in Euclidean space, but it is imaginary in Lorentzian signature. Thus, for the case $n=3$ the Wick rotation is potentially obstructed by the presence of the Neveu-Schwarz $B$-field in heterotic string theory.\footnote{Another example of this type are $AdS\times S$ vacua of type II string theory. In~\cite{Balasubramanian:2001rb}, it was proposed that these backgrounds can be interpreted as Euclidean de Sitter solutions, but this interpretation leads to imaginary RR form fields.} However, for $n>3$, e.g. for the case of $dS_4$, we do not expect such an obstruction to exist. At the sigma model level, this is the statement that there are no non-trivial $p<n$ form field strengths on an $n$-dimensional de Sitter space that preserve the full $SO(1,n)$ symmetry. This is derived in Appendix~\ref{invariantforms}.  Of course, if the de Sitter vacuum is inherently stringy, the sigma model may not provide a quantitatively accurate description, but we are assuming that it provides a good qualitative guide in the above sense. 
\end{enumerate}

\noindent
Under these assumptions, the Wick rotated background  is a compact unitary CFT with a global $SO(n+1)$ symmetry realized as in \C{ttww}. As is well known,  this means that the currents $J^a$ and $\bar J^a$ are separately conserved, $\bar\partial J^a=0$, $\partial\bar J^a=0$. Indeed, conformal invariance implies that the two point function of the currents is given by 
 \begin{align}
 \label{thr}
 \langle J^a(z)J^b(0)\rangle = {k\over z^2}~,\qquad  \langle \bar J^a(\bar z)\bar J^b(0)\rangle = {\bar k\over \bar z^2}~.
\end{align}
This means that the two point function $\langle \bar\partial J^a(z)\bar\partial J^b(0)\rangle=0$ for separated points, so the operator $\bar\partial J^a$ is null. In a unitary compact CFT such an operator must vanish. Similarly, one has $\partial\bar J^a=0$.

The rotation generators in CFT on $\R^n$, $J^{ij}=x^{[i}\partial x^{j]}$ and  $\bar J^{ij}=x^{[i}\bar\partial x^{j]}$, are an example for which \C{ttww}\ is valid, but \C{thr}\ is not. This is because of the non-compactness of the CFT, and in particular the presence of a continuum of states starting at $L_0=0$. This phenomenon cannot occur in a theory with a discrete spectrum.

The fact that the left and right-moving currents are separately conserved implies that the theory is actually invariant under two copies of an $SO(n+1)$ symmetry, which is more than one would expect. One might think that a natural solution for this issue is that one of the two copies vanishes. For example, if the level of the left-moving current algebra, $k$, vanishes, the operators $J^a$ must vanish as well, and there is only one set of $SO(n+1)$ currents, $\bar J^a$, as expected. However, this possibility suffers from the problem that the resulting theory does not contain a gravity sector. 

Indeed, based on experience from flat space-time, and the world-sheet construction of string theory on group manifolds (such as $AdS_3$ and $S^3$), we expect the gravity sector on a de Sitter background to be described by vertex operators of the form $\xi_{ab}J^a\bar J^b V$, where $V$ is a vertex operator that transforms as a primary of the affine Lie algebra, and $\xi_{ab}$ a polarization tensor. If either $J^a$ or $\bar J^a$ are absent, such vertex operators do not exist, and it is not clear in what sense one can interpret the resulting construction as describing a stringy generalization of gravity on $dS_n$. In any case, we will assume that the theory contains gravitons, which means that $k$ and $\bar k$ are both strictly positive. 

In fact, the constraint on the left moving level, $k$, is stronger since it is the total level of a super Kac-Moody algebra.\footnote{A pedagogical discussion of super Kac-Moody algebras may be found in chapter 18 of~\cite{Polchinski:1998rr}.}    Recall that a level $k$ super Kac-Moody algebra consists of a bosonic KM algebra with level $k_B$ and free fermions that transform in the adjoint of $G$.  The latter generate a Kac-Moody algebra with level given by $h_G$, the dual Coxeter number of $G$.  The total level is then $k=k_B +h_G$. Unitarity requires $k_B$ to be a non-negative integer. For $G=SO(n+1)$ one has $h_G=n-1$. Thus, the left-moving level satisfies the constraint 
\be
 k\ge n-1~.
 \ee
There is also an upper bound on $k$ from unitarity. In any theory with a super Kac-Moody symmetry one can write the central charge as a sum of two contributions, that of the super KM sector, 
\be
 c_{SKM}=\left({k_B\over k} +{1\over 2}\right)\dim G
 \ee
 and that of the coset obtained by modding out the CFT by the super KM algebra, $c_{\rm coset}$. If the original CFT is unitary, so is the coset, which means that $c_{\rm coset}\ge 0$. Since the total central charge is $c_{SKM}+c_{\rm coset}=15$, and $\dim SO(n+1)=n(n+1)/2$, we get a constraint on the level $k$,
\be
 \label{ffrr}
 n-1\le k \le {n-1\over {3\over2}-{30\over n(n+1)}}~.
 \ee
For $n\ge 8$ no solutions exist. For $n=7$ there is a unique solution, $k=6$, which corresponds to $c_{\rm coset}=1$, i.e. the coset is an $N=1$ minimal model corresponding to a scalar field on a circle of a particular radius, and the full solution is a product of an $N=1$ supersymmetric WZW model with $G=SO(8)$ and the minimal model. In fact, the bosonic level $k_B=0$ for this case, so the supersymmetric WZW model is a theory of $\dim SO(8)=28$ free fermions transforming in the adjoint representation. 

This model does not seem to have an interpretation as a Euclidean $dS_7$ background, despite its $SO(8)$ symmetry. In particular, it is not clear that there is a continuation to Lorentzian signature, with $SO(1,7)$ symmetry and a sensible semi-classical interpretation. It is not surprising that this is the case -- theories with Kac-Moody symmetry correspond to sigma models on group manifolds and are inherently Euclidean.

For lower $n$, one finds two solutions with $n=6$, corresponding to $k_B=0$, $c_{\rm coset}=9/2$ and $k_B=1$  and $c_{\rm coset}=1$. For $n=5$, there are five solutions \begin{align}
(k_B,c_{\rm coset}) \in \{(0,15/2), (1,9/2),  (2,5/2), (3,15/14),  (4,0)\}~.
\end{align}
For $n=4$ there is no upper bound on $k$ from~\C{ffrr}, but there is still just a small number of possible values of $(k_B,c_{\rm coset})$.  For $k_B >17$ we find that $c_{\rm coset} < 3/2$, which means the coset CFT must be a unitary N=1 minimal model, and therefore
\begin{align}
c_{\rm coset} = \frac{30}{k_B+3}=\frac{3}{2} \left( 1-\frac{8}{p(p+2)} \right)
\end{align}
for some integer $p> 2$.  The only solutions are $(k_B,p) \in \{(18,12), (21,6), (27,4) \}$. 

All the above solutions suffer from two problems:  they exist only for a few finite values of $k$, so the size of the would be de Sitter space, which goes like $L^2\sim kl_s^2$, is bounded;\footnote{A simple way to understand this scaling is to note that the world-sheet dimensions of primaries go like $C_2(R)/k$, where $C_2$ is the quadratic Casimir in the representation $R$.}   even more seriously, they do not seem to have the interpretation of corresponding to a de Sitter space, and in particular do not have a natural continuation to Lorentzian signature. 

To summarize, we see that the presence of separately conserved left and right-moving currents in potential world-sheet constructions of $dS_n$ space-times with $n\ge 4$ essentially rules out the possibility that such spaces arise in heterotic string theory.  Note that we do not require that the string dilaton be stabilized, or even that the world-sheet theory be free of tachyons. There are no (macroscopic) stable or unstable de Sitter solutions at tree-level in string perturbation theory. 

It is worth mentioning one possible way to evade this constraint. If one could find a background in which the world-sheet ghost sector does not decouple from the matter sector then unitarity of the world-sheet theory would no longer be guaranteed, invalidating one of the requirements for promoting a space-time symmetry to a Kac-Moody symmetry.  For type II string theories, it is easy to find such backgrounds by turning on RR fluxes. In the case of the heterotic string, we are not aware of any simple background that requires a mixing of ghost and matter sectors. It would be very interesting to find other ways to evade our constraint. 

Finally, we note that our argument does not generalize to $AdS_n$.  As shown in the Appendix, we can still argue that there is no obstruction to the Wick rotation for $n>3$, but the resulting Euclidean theory is not compact, and we cannot immediately conclude that the $AdS_n$ isometries must be realized by KM symmetries. This is an interesting direction for future study.

\section*{Acknowledgements}

It is our pleasure to thank Michael Douglas, Juan Maldacena, Emil Martinec and Callum Quigley for helpful discussions.  S.~S. would like to thank the organizers of the Madrid ``Fine-Tuning, Anthropics and the String Landscape" workshop, and the organizers of the PCTS workshop on ``String Cosmology and Inflation" for their hospitality.  We would also like to thank the Mitchell Institute for hospitality during the workshop on ``Heterotic Strings and (0,2) QFT.''  DK is supported in part by DOE grant DE FG02-13ER41958. IVM is supported in part by NSF Focused Research Grant DMS-1159412. TM and SS are supported in part by NSF Grant No.~PHY-1316960.

\appendix
\section{Invariant Forms on $(A)dS$} \label{invariantforms}
In this appendix we classify isometry-invariant forms on $(A)dS$ space.  We show that the only such forms are constants and constant multiples of the volume form on those space-times.

Let $X =\R^{p,q}$ and let $\eta$ to be the canonical flat metric with signature $(p,q)$.  Fixing a global set of coordinates $x^a$, $a=1,\ldots, n+1$ on $X$, set 
\begin{align}
\rho = x^a x^b \eta_{ab}~.
\end{align}
Let $M$ be a connected component of the vanishing set of
\begin{align}
f(x) = \rho+ t
\end{align}
for some non-zero constant $t$.  By pulling back the flat metric $\eta$ to the hypersurface $M$ we obtain our $(A)dS_n$ space.  The $\frac{1}{2} n(n+1)$ Killing vectors on $M$ are obtained by projecting the Lorentz generators on $X$.    These have the form
\begin{align}
V = x^a \eta_{ab} \Lambda^{bc} \frac{\p}{\p x^c}~,
\end{align}
where $\Lambda$ is a constant anti-symmetric matrix, and since $\cL_V f = \cL_V \rho =0$, every $V$ preserves  the hypersurface for any $t$.

Consider a small tubular neighborhood  $N$ of $M$ in $X$ parametrized by $\rho$ and local coordinates on $M$.  $N$ is in fact diffeomorphic to the total space of the normal bundle to $M$ in $X$, $\pi: N \to M$.  If $\omega_0 \in \Omega^k(M)$ is an isometry-invariant form, then $\pi^\ast\omega_0$ is an isometry-invariant form on $N$, and multiplying this by a suitable bump function $b (\rho)$, we obtain a Lorentz-invariant $k$-form $\omega$ on $\R^{p,q}$ whose pull-back to $M$ yields $\omega_0$.

Lorentz-invariant forms on $\R^{p,q}$ are familiar from basic representation theory.  Up to multiplying by a function of $\rho$  the invariants are: $d\rho$ and $\dVol(X)$, which pull-back to $0$ on $M$; and $1$ and $\ast_{\eta} d\rho = \ep_{a_1\cdots a_{n+1}} x^{a_1} dx^{a_2} \cdots dx^{a_{n+1}}$, which pull-back to $1$ and $\dVol(M)$.


\ifx\undefined\bysame
\newcommand{\bysame}{\leavevmode\hbox to3em{\hrulefill}\,}
\fi

\end{document}